\documentclass[12pt]{article}
\usepackage{epsf}
\usepackage{amsmath,amssymb}
\usepackage{graphicx}
\usepackage{color}
\usepackage{here}
\usepackage{cite}
\usepackage[colorlinks,citecolor=blue]{hyperref}
\renewcommand{\thefootnote}{\#\arabic{footnote}}

\renewcommand{\thefootnote}{\fnsymbol{footnote}}
\def\thefootnote{\fnsymbol{footnote}}

\makeatletter

\@addtoreset{equation}{section}
\makeatother

\begin{document}

\begin{titlepage}

\begin{center}

\vskip .75in

{\Large \bf Impact of big bang nucleosynthesis on the $H_0$  tension}  

\bigskip

\vskip .75in

{\large
Tomo~Takahashi$\,^1$ and Yo~Toda$\,^2$ 
}

\vskip 0.25in

{\em
$^{1}$Department of Physics, Saga University, Saga 840-8502, Japan  \vspace{2mm} \\
$^{2}$Department of Physics, Hokkaido University, Sapporo 060-0810, Japan 
}

\end{center}
\vskip .5in

\begin{abstract}

We investigate the impact of big bang nucleosynthesis (BBN) on the Hubble tension, focusing on how the treatment of the reaction rate and observational data affect the evaluation of the tension. We show that  the significance of the tension can vary by $0.8 \sigma$ in some early dark energy model, depending on the treatment of the reaction rate and observational data. This indicates that how we include the BBN data in the analysis can give a significant impact on the Hubble tension, and we need to carefully consider the assumptions of the analysis to evaluate the significance of the tension when the BBN data is used.

\end{abstract}

EPHOU-23-011

\end{titlepage}

\renewcommand{\thepage}{\arabic{page}}
\setcounter{page}{1}
\renewcommand{\thefootnote}{\#\arabic{footnote}}
\setcounter{footnote}{0}

\section{Introduction \label{sec:intro}}

The so-called $\Lambda$CDM model has now been recognized as the standard paradigm of cosmology.  Indeed almost all observations can be understood in the framework of the $\Lambda$CDM model, however, there have been several tensions reported,  among which the Hubble tension ($H_0$ tension) is one of the most pronounced one where the values of the Hubble constant $H_0$ are inconsistent at almost 5$\sigma$ level between the ones obtained from direct measurements such as Cepheid calibrated distance ladder $H_0 = 73.30 \pm 1.04~{\rm km/s/Mpc}$ \cite{Riess:2021jrx}\footnote{
Here we quote the value based on the analysis including high-redshift supernovae. 
} and from indirect observations such as cosmic microwave background (CMB) from Planck $H_0 = 67.66 \pm 0.42~{\rm km/s/Mpc}$ \cite{Planck:2018vyg}.  Even if different data are used, local direct observations consistently infer a relatively higher value of $H_0$ compared to the one obtained from  Planck, on the other hand, indirect observations, even without Planck data, gives $H_0$ consistent with that obtained from Planck\footnote{
Even without CMB data, the combination of data from baryon acoustic oscillation, type Ia supernovea and big bang nucleosynthesis also gives a relatively low value of $H_0$ in the framework of $\Lambda$CDM and other extended models \cite{Schoneberg:2019wmt,Okamatsu:2021jil,Schoneberg:2022ggi}. 
}  (see, e.g., \cite{DiValentino:2021izs,Perivolaropoulos:2021jda} for reviews of the $H_0$ tension).  

A lot of works have been devoted to resolving the $H_0$ tension by extending the $\Lambda$CDM framework  (for various models proposed to solve the tension, see, e.g., reviews \cite{DiValentino:2021izs,Schoneberg:2021qvd}). In those models, the value of $H_0$ can become higher than $\Lambda$CDM  case when they are fitted to CMB in combination with some other data such as baryon acoustic oscillation (BAO), type Ia supernova (SNeIa), and so on. It would be important to notice that there exists the correlation between $H_0$ and other cosmological parameters in the fit to CMB\footnote{
One can easily recognize such a correlation from 2D constraints in the plane of $H_0$ and some other parameters (e.g., \cite{Planck:2018vyg}).  See also, e.g., \cite{Ichikawa:2004zi,Ichikawa:2008pz,Sekiguchi:2009zs} to see how $H_0$ is correlated with other cosmological parameters in the structure of acoustic peaks in CMB power spectrum. 
}.  Although such a correlation is somewhat involved in models proposed to resolve the $H_0$ tension,  the increase of the value of $H_0$ certainly affects other aspects of cosmology. Indeed the baryon energy density generally tends to become larger than that for the case with the $\Lambda$CDM in models proposed to resolve the $H_0$ tension when fitted to CMB in combination with other observations \cite{Takahashi:2022cpn,Seto:2021xua,Seto:2022xgx,Seto:2021tad}\footnote{
It has also been argued that constraints on the neutrino mass \cite{Sekiguchi:2020igz}, the spectral index for primordial power spectrum $n_s$ \cite{Ye:2021nej,Takahashi:2021bti,Jiang:2022uyg,Ye:2022efx} are also modified in models proposed to resolve the tension such as time-varying electron mass \cite{Sekiguchi:2020teg} and  early dark energy models \cite{Poulin:2018cxd}.
}. 

The baryon density can also be well determined by big bang nucleosynthesis (BBN) and hence we can also include BBN data in the fitting to derive the cosmological parameters. However, as we discuss in this paper, we need to be cautious regarding what observational data we use and what nuclear reaction rates are adopted, depending on which the derived values of the cosmological parameters can change. Therefore when the BBN data is combined with CMB and some other data, it can affect the value of $H_0$ in models to resolve the $H_0$ tension through the correlations among the cosmological parameters in the fitting. Since, among the light elements, the baryon density is mainly determined by the primordial deuterium abundance, which we denote $D_p$ in the following,  we mainly investigate the impact of the treatment of the primordial deuterium on the $H_0$ tension\footnote{
The primordial helium abundance $Y_p$ is also precisely measured \cite{Aver:2015iza,Hsyu:2020uqb,Kurichin:2021ppm,Matsumoto:2022tlr},  however $Y_p$ is not so severely constrained by CMB data  (see, e.g., \cite{Planck:2018vyg,Ichikawa:2006dt,Ichikawa:2007js}), and hence the inclusion of $Y_p$ data would not affect  $H_0$ much. Therefore we focus on the treatment of the deuterium abundance in this paper although the $Y_p$ data of \cite{Aver:2015iza} 
is included in our analysis.
}.  To this end, we consider several different treatments to include the primordial deuterium abundance in the analysis:  For observational data, we use two different data for $D_p$ from Cooke et al. \cite{Cooke:2017cwo} and the weighted mean of the recent 11 measurements compiled by Particle Data Group \cite{ParticleDataGroup:2022pth}. For the theoretical calculations of $D_p$, we adopt the nuclear reaction rate for $d (p, \gamma) ^3 {\rm He}$ obtained from the extrapolation of experiments \cite{Adelberger:2010qa,Pisanti:2020efz} and the theoretically evaluated one via an ab-initio approach \cite{Marcucci:2015yla}.  As we argue in this paper, the significance of the $H_0$ tension can vary up to 0.8$\sigma$ depending on  the treatment of how we include $D_p$ in the analysis and models assumed,  which shows that BBN has a significant impact on the $H_0$ tension. 

The organization of this paper is as follows. In the next section, we summarize our analysis method and models assumed in the analysis. Then, in Section~\ref{sec:results}, we show our results and argue that the treatment of BBN in the analysis can give a significant impact on the $H_0$ tension depending on what observational data and  the reaction rate  are adopted. We give the summary of this paper in the final section.


\section{Models and analysis method \label{sec:setup}}


\subsection{Models}

To investigate the impact of the treatment of the primordial deuterium abundance 
on the $H_{0}$ tension, we consider two model frameworks: a flat
$\Lambda$CDM and early dark energy (EDE) models \cite{Poulin:2018cxd}.
In the framework of a flat $\Lambda$CDM model, we vary 6 standard
cosmological parameters: baryon density~$\Omega_{b}h^{2}$, cold dark
matter (CDM) density~$\Omega_{c}h^{2}$, the acoustic scale~$\theta$,
the reionization optical depth~$\tau$, the amplitude of primordial
power spectrum~$A_{s}$ and its spectral index~$n_{s}$. Since we
assume a flat Universe, the Hubble constant $H_{0}$ is derived once
these parameters are given. The amplitude $A_{s}$ is defined at the
reference scale $k_{{\rm ref}}=0.05~{\rm Mpc}^{-1}$.

EDE model \cite{Poulin:2018dzj,Poulin:2018cxd,Braglia:2020bym,Agrawal:2019lmo} has been extensively
studied as a possible candidate to resolve the $H_{0}$ tension (for
recent review on early dark energy, see \cite{Kamionkowski:2022pkx,Poulin:2023lkg}).
There are several variations of EDE model discussed
in the literature depending on the potential of a scalar field and/or
the actual implementation of the model, here we adopt the one based on an axionlike field \cite{Poulin:2018dzj}\footnote{
The cosmic birefringence suggested by the recent analyses of the Planck data \cite{Minami:2020odp,Diego-Palazuelos:2022dsq} 
also motivates to introduce such an axionlike field $\phi$ \cite{Capparelli:2019rtn,Murai:2022zur}.
} whose potential is assumed as 
\begin{equation}
V (\phi) = \Lambda^4 \left( 1 - \cos (\phi / f) \right)^n \,,
\end{equation}
where $\Lambda$ sets the energy scale of the potential and $f$ is a  parameter which could be related to a breaking scale of the model.  To describe the energy fraction of the EDE field at the redshift when $\phi$ starts to oscillate, denoted as $z_c$, one usually introduces the following quantity:
\begin{equation}
f_{\rm de} (z_c) = \frac{\rho_{\rm de} (z_c)}{\rho_{\rm total}  (z_c)} \,,
\end{equation}
where $\rho_{\rm de}$ and $\rho_{\rm total}$ are energy densities of the EDE field and the total component, respectively. When $n=2$, the energy density of EDE scales as $\rho_{\rm de} \propto a^{-4}$.  When $n=\infty$, $\rho_{\rm de}$ decreases as $\rho_{\rm de} \propto a^{-6}$.
We use the \texttt{camb} \cite{camb} where axionlike EDE is already implemented\footnote{
The fluid approximation as in Eq.~(16) in the paper~\cite{Poulin:2018dzj} is used for an axionlike field,  which is implemented in the background and perturbation equations.
}
and perform the MCMC analysis sampling $f_{\rm de} (z_c)\in[0.00001,0.15]$, $z_c\in[1000,50000]$, and $\Theta_i\equiv \phi_{\rm ini} / f \in[0.01,3.14]$ (the initial value of $\phi$) in addition to the 6 standard parameters.

\subsection{Analysis method}

To evaluate the impact of the treatment of BBN on the Hubble tension,
we perform Markov Chain Monte Carlo analysis by using a modified version
of \texttt{CosmoMC} \cite{Lewis:2002ah}, which accommodates the EDE model discussed in \cite{Poulin:2018dzj}, by using the
data from Planck (TT,TE,EE+lowE)~\cite{Planck:2019nip} and BAO \cite{Beutler:2011hx,Ross:2014qpa,BOSS:2016wmc}.
In some cases, we assume a prior on the Hubble constant as $H_{0}=73.30\pm1.04$
\cite{Riess:2021jrx} (hereinafter, we refer this prior as R21,  after the author and the year of the paper). In addition to these, we include the light
element data of the primordial helium abundance $Y_{P}=0.2449\pm0.0040$ \cite{Aver:2015iza}
and two different
data for the primordial deuterium abundance from Cooke et al. $D_{p}=(2.527\pm0.030)\times10^{-5}$
\cite{Cooke:2017cwo} and the weighted mean of the recent 11 measurements
compiled by Particle Data Group $D_{p}=(2.547\pm0.025)\times10^{-5}$
\cite{ParticleDataGroup:2022pth}. 

For the theoretical calculations of light elements,
we consider three different nuclear rate for $d(p,\gamma)^{3}{\rm He}$,
which is important in computing the deuterium abundance. The one is from
Adelberger et al. \cite{Adelberger:2010qa} which is evaluated from
the extrapolation of experimental results.
Another reaction rate is from Marcucci
et al. \cite{Marcucci:2015yla} which is an ab-initio theoretical
calculation of the reaction rate,  which shows a larger $d(p,\gamma)^{3}{\rm He}$ rate, more deuterium burning, and less abundance of deuterium $D_p$.
The other is taken from Pisanti et al.~\cite{Pisanti:2020efz} 
which is the extrapolation of the most recent experimental results and their value is somewhat in between the two  rates mentioned above. 
The value of the theoretical uncertainties are given by 
$\sigma_{Y_{p},\rm th}=0.0003$ \cite{Consiglio:2017pot} for the all analysis,  $\sigma_{D_{p},\rm th}=0.06 \times10^{-5}$  \cite{Adelberger:2010qa} for the analysis using Adelberger et al.,  $\sigma_{D_{p},\rm th}= 0.03 \times10^{-5}$ \cite{Marcucci:2015yla} for the analysis using Marcucci et al., and $\sigma_{D_{p},\rm th}= 0.06 \times10^{-5}$~\cite{Gariazzo:2021iiu} 
for the analysis using Pisanti et al. In total, we perform 7 different
analyses depending on the assumption for the data and the theoretical
calculation of $D_{p}$, which is summarized in Table~\ref{tab:Dp_treatment}.
``offset" in  Table~\ref{tab:Dp_treatment} indicates that we adopt the theoretical bias offset
$-0.091 \times 10^{-5}$ and corrected theoretical error $\sigma_{D_{p},\rm th}=0.089 \times10^{-5}$.
These values are introduced in the standard \texttt{CosmoMC} analysis 
so that the BBN constraints correspond to the prior  $\Omega_b h^2 = 0.0222\pm 0.005$\footnote{
See Sec. 7.6.1 of the Planck paper~\cite{Planck:2018vyg} for details.
}. We note that Case II or III is commonly adopted one in the literature when the BBN data is included in the CMB analysis.

\begin{table}
\centering %
\begin{tabular}{l|cc}
\hline 
 & Reaction rate for $d(p,\gamma)^{3}{\rm He}$  & $D_{p}$ data \tabularnewline
\hline \hline
{ } Case I  & Adelberger et al.  & no-data\tabularnewline
{ } Case II  & Adelberger et al.  & Cooke et al. + offset \tabularnewline
{ } Case III  & Adelberger et al.  & Cooke et al.\tabularnewline
{ } Case IV  & Marcucci et al.  & Cooke et al.\tabularnewline
{ } Case V & Marcucci et al.  & recent 11 weighted mean\tabularnewline
{ } Case VI & Pisanti et al. & Cooke et al.\tabularnewline
{ } Case VII & Pisanti et al. & recent 11 weighted mean\tabularnewline
\hline \hline
\end{tabular}\caption{\label{tab:Dp_treatment} 
Combinations of the data of $D_{p}$ and theoretical treatment
of the reaction rate for $d(p,\gamma)^{3}{\rm He}$ used in our analysis. }
\end{table}


\section{Results \label{sec:results}}


Now we present  the results of our analysis. In Tables~\ref{tab:LCDM},
\ref{tab:EDE_n2} and \ref{tab:EDE_ninf}, the derived values of $H_{0}$
and $\Omega_{b}h^{2}$, as well as the Gaussian tension are summarized
for the framework models of $\Lambda$CDM, EDE ($n=2$) and EDE ($n=\infty$),
respectively. The Gaussian tension is evaluated between the direct
measurement of $H_{0}$ from Cepheid calibrated supernova distance
ladder \cite{Riess:2021jrx} $H_{0}=73.30\pm1.04~{\rm km/s/Mpc}$
and the one obtained in our analysis using CMB+BAO (with/without
BBN and the $H_{0}$ prior). For the cases of EDE, we also give the
value of $f_{{\rm de}}(z_{c})$. We have performed 10 analyses including
Case I - VII in Table~\ref{tab:Dp_treatment} and the cases with and without the prior on $H_{0}$ (R21) for some cases.

\begin{figure}
\begin{center}
\includegraphics[width=10cm]{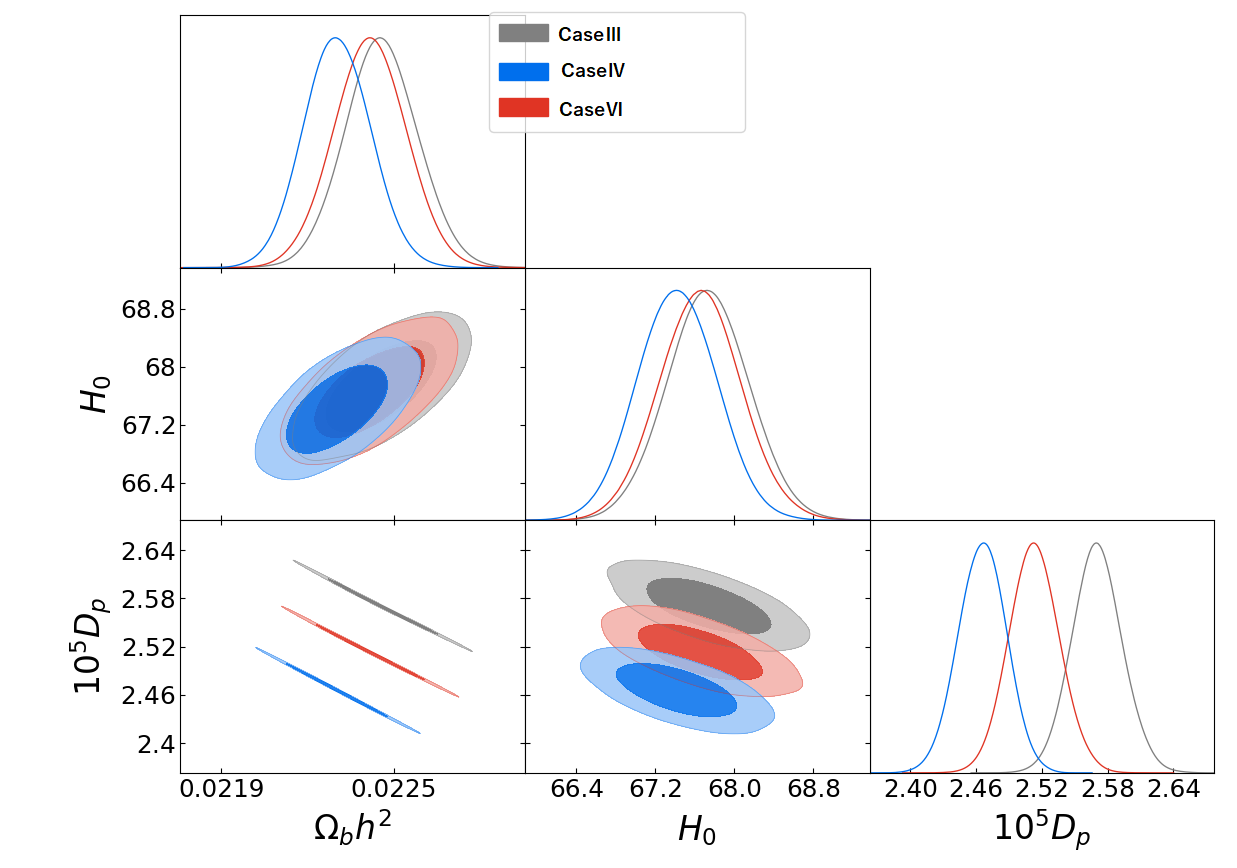}
\end{center}
\caption{\label{tab:LCDM_fig} Posterior distributions and two dimensional constraints for $\Omega_{b}h^2, H_0$ and $D_p$ in the $\Lambda$CDM model. }
\end{figure}

For the case of $\Lambda$CDM, depending on the treatment of the observational
data and theoretical calculation for $D_{p}$, the significance of
the tension varies from $4.97\sigma$ to $5.38\sigma$ when R21 is not included. As already mentioned, the setup
of Case II or III is commonly adopted when one includes the BBN data in combination
with CMB. It is also worth noting that in Case III the Hubble tension is slightly eased 
with respect to Case I (no-BBN data analysis) and Case II (standard BBN analysis including the offset).
In any case, just by changing the assumption on the observational
data for $D_{p}$ and the reaction rate, the significance is increased
by $0.41\sigma$. The difference mainly comes from the treatment of
the reaction rate. In Figure~\ref{tab:LCDM_fig}, we show the posterior distributions and two dimensional constraints for $\Omega_{b}h^2, H_0$ and $D_p$, which shows that 
the deuterium abundance $D_p$ decrease 
in the order of Case~III (Adelberger et al.), Case~VI (Pisanti et al.), and Case~IV (Marcucci et al.) for a given baryon fraction $\Omega_{b}h^2$ (all these cases use Cooke et al. \cite{Cooke:2017cwo} for the observational data of $D_p$).
In order to be consistent with $D_p$ observations, it is important to notice that the decrease of $D_p$ leads to a lower baryon fraction, a larger sound horizon $r_{*}$, and a lower Hubble constant $H_0$, which makes the tension worse. 
The observational data also affects slightly
although two data from Cooke et al. \cite{Cooke:2017cwo} and the
averaged mean value from 11 recent measurements of $D_{p}$ \cite{ParticleDataGroup:2022pth}
are almost consistent. For example, by comparing Case~IV  and V in which Cooke et al. and the recent 11 weighted mean value are adopted for the $D_p$ data with the same reaction rate, the tension in Case~V is slightly larger than that in  Case~IV. 

\begin{table}
\centering %
\begin{tabular}{l|ccc}
\hline 
 & $\Omega_{b}h^{2}$  & $H_{0}$  & Tension \tabularnewline
\hline 
\hline 
{ } 
Case I  & $0.02242\pm0.00014$  & $67.67\pm0.43$  & 5.00$\sigma$ \tabularnewline
{ } 
Case I +R21  & $0.02259\pm0.00013$ & $68.49\pm0.40$ & 4.32$\sigma$ \tabularnewline
{ } 
Case II & $0.02240\pm0.00013$  & $67.63\pm0.43$  & 5.04$\sigma$ \tabularnewline
{ } 
Case III & $0.02246\pm0.00013$  & $67.73\pm0.42$  & 4.97$\sigma$ \tabularnewline
{ } 
Case III+R21 & $0.02260\pm0.00012$  & $68.51\pm0.39$  & 4.31$\sigma$\tabularnewline
{ } 
Case IV  & $0.02230\pm0.00012$  & $67.42\pm0.40$  & 5.28$\sigma$ \tabularnewline
{ } 
Case V  & $0.02226\pm0.00011$  & $67.33\pm0.39$  & 5.38$\sigma$ \tabularnewline
{ } 
Case V +R21 & $0.02238\pm0.00011$ & $68.09\pm0.38$ & 4.71$\sigma$\tabularnewline
{ } 
Case VI  & $0.02242\pm0.00013$ & $67.66\pm0.41$ & 5.05$\sigma$\tabularnewline
{ } 
Case VII  & $0.02240\pm0.00013$ & $67.62\pm0.42$ & 5.06$\sigma$\tabularnewline
\hline \hline
\end{tabular}\caption{ \label{tab:LCDM} Derived values of $\Omega_{b}h^{2}$,  $H_{0}$ and the Gaussian tension for $\Lambda$CDM model. }
\end{table}

\begin{figure}
\begin{center}
\includegraphics[width=10cm]{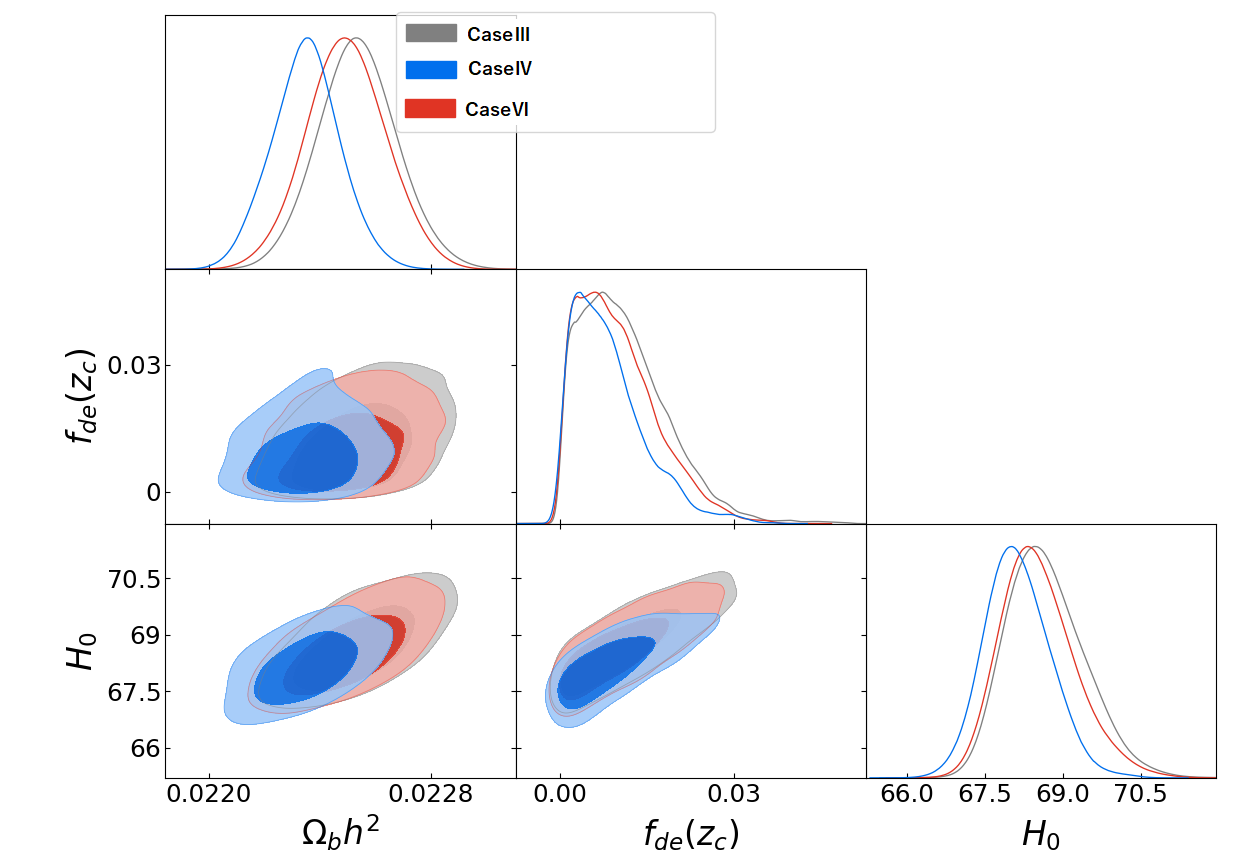}
\end{center}
\caption{\label{fig:EDEn2_fig}  Posterior distributions and two dimensional constraints for $\Omega_{b}h^2, f_{\rm DE} (z_c)$ and $H_0$ for the EDE model with $n=2$.}
\end{figure}

\begin{figure}
\begin{center}
\includegraphics[width=10cm]{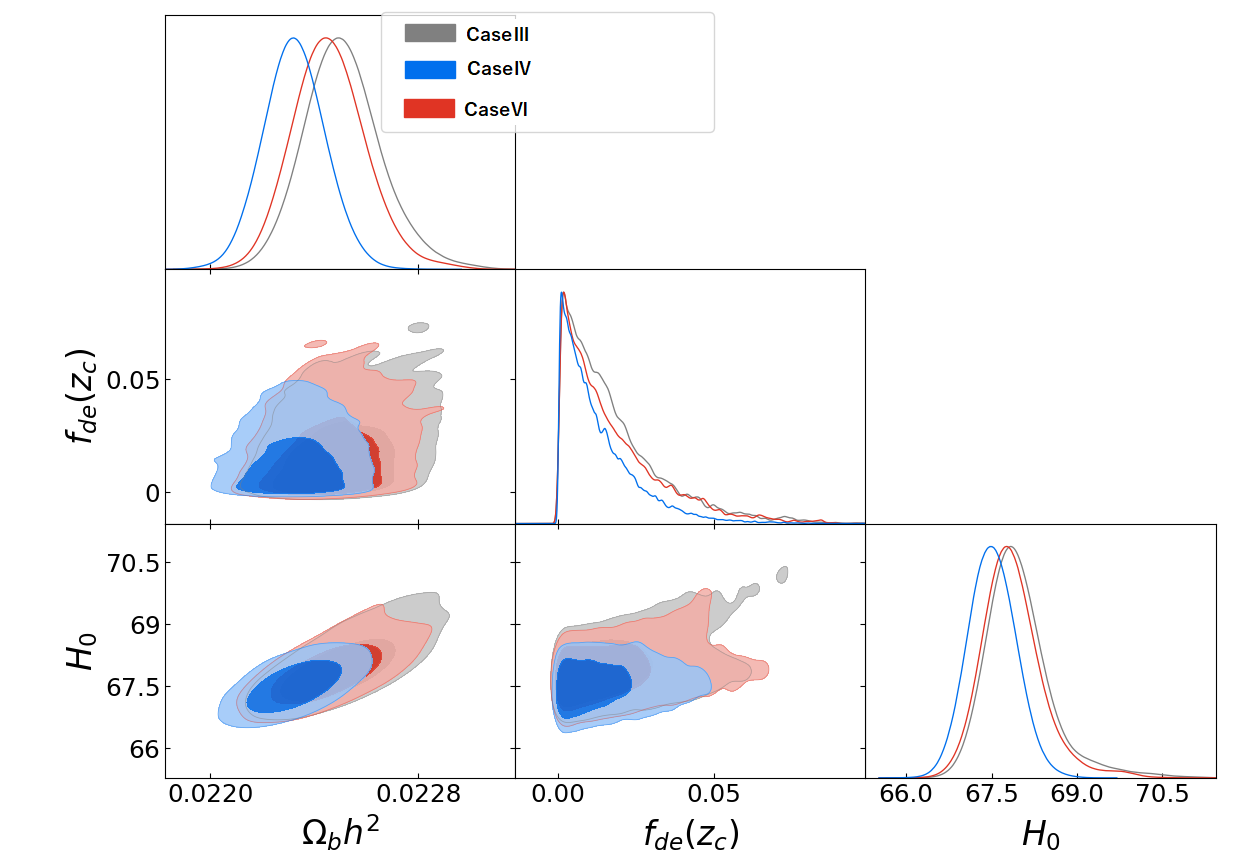}
\end{center}
\caption{\label{fig:EDEninf_fig} Posterior distributions and two dimensional constraints for $\Omega_{b}h^2, f_{\rm DE} (z_c)$ and $H_0$ for the EDE model with $n=\infty$.}
\end{figure}

The impact of the treatment of BBN is more significant when we study
the issue in the EDE model which has been proposed to relax the $H_0$ tension. In Tables~\ref{tab:EDE_n2} and \ref{tab:EDE_ninf}, derived values of $\Omega_{b}h^{2},f_{{\rm de}}(z_{c})$, 
 $H_{0}$ and the Gaussian tension are listed for the cases of $n=2$ and $n=\infty$. In Figs.~\ref{fig:EDEn2_fig} and \ref{fig:EDEninf_fig}, the posterior distribution and two dimensional constraints for $\Omega_{b}h^2, f_{\rm DE} (z_c) $ and $H_0$ are depicted for the cases of $n=2$ and $n=\infty$, respectively. 
We first  note that, in the EDE model we adopted here, the significance of the tension is
$3.6\sigma$ for Case~I where we do not include the BBN data, which can be compared to $5.0\sigma$ in the counterpart of the $\Lambda$CDM.
However, when the BBN is included
in the analysis for the case of $n=2$, it increases to $3.83\sigma$
for the standard treatment of Case~II and even worse to $4.41\sigma$
for Case~V. It should be noticed that the significance of the tension
varies from $3.61\sigma$ to $4.41\sigma$ ($0.8\sigma$ difference)  just by changing the assumption
on the treatment of $D_{p}$. The tendency of which reaction rate makes the tension larger is the same as that in the $\Lambda$CDM model, however, the change of the tension is more pronounced in the EDE model. 
For the case of $n=\infty$, the Gaussian
tension is increased from $4.58\sigma$ in Case~II to $5.23\sigma$
in Case~V, which is $0.65\sigma$ increase. Although the change is less significant compared to the
case of $n=2$, BBN still have an impact on the $H_{0}$
tension.

\noindent 
\begin{table}
\centering %
\begin{tabular}{l|cccc}
\hline 
 & $\Omega_{b}h^{2}$  & $f_{{\rm de}}(z_{c})$  & $H_{0}$  & Tension \tabularnewline
\hline 
\hline 
{ } Case I  & $0.02252\pm0.00016$  & $0.0106_{-0.0088}^{+0.0043}$  & $68.64_{-0.86}^{+0.65}$  & $3.61\sigma$ \tabularnewline
{ } 
Case I +R21  & $0.02277\pm0.00016$  & $0.0277_{-0.0092}^{+0.0071}$  & $70.64\pm0.73$  & $2.09\sigma$ \tabularnewline
{ } 
Case II & $0.02245\pm0.00015$  & $0.0098_{-0.0082}^{+0.0039}$  & $68.49_{-0.79}^{+0.60}$  & $3.83\sigma$ \tabularnewline
{ } 
Case III & $0.02254\pm0.00014$  & $0.0111_{-0.010}^{+0.0036}$  & $68.68_{-0.86}^{+0.61}$  & $3.61\sigma$ \tabularnewline
{ } 
Case III+R21 & $0.02276_{-0.00014}^{+0.00016}$  & $0.0261_{-0.0081}^{+0.0073}$  & $70.54\pm0.70$  & $2.20\sigma$ \tabularnewline
{ } 
Case IV  & $0.02235\pm0.00012$  & $0.0085_{-0.0078}^{+0.0028}$  & $68.13_{-0.68}^{+0.57}$  & $4.26\sigma$ \tabularnewline
{ } 
Case V  & $0.02229\pm0.00012$  & $0.0081_{-0.0079}^{+0.0021}$  & $67.96_{-0.72}^{+0.50}$  & $4.41\sigma$ \tabularnewline
{ } 
Case V +R21 & $0.02238\pm0.00015$  & $0.033_{-0.015}^{+0.012}$  & $70.22_{-0.88}^{+0.76}$  & $2.32\sigma$ \tabularnewline
{ } 
Case VI  & $0.02250\pm0.00014$  & $0.0099_{-0.0089}^{+0.0034}$  & $68.53_{-0.81}^{+0.58}$  & $3.80\sigma$ \tabularnewline
{ } 
Case VII  & $0.02247\pm0.00014$  & $0.0096_{-0.0086}^{+0.0032}$  & $68.46_{-0.80}^{+0.58}$  & $3.86\sigma$ \tabularnewline
\hline \hline
\end{tabular}

\caption{ \label{tab:EDE_n2} Derived values of $\Omega_{b}h^{2},f_{{\rm de}}(z_{c})$, 
 $H_{0}$  and the Gaussian tension  for the EDE model with $n=2$. }
\end{table}

\begin{table}
\centering %
\begin{tabular}{l|cccc}
\hline 
 & $\Omega_{b}h^{2}$  & $f_{{\rm de}}(z_{c})$  & $H_{0}$  & Tension \tabularnewline
\hline 
\hline 
{ } Case I  & $0.02249_{-0.00017}^{+0.00014}$  & $0.0174_{-0.017}^{+0.0037}$  & $67.93_{-0.62}^{+0.42}$  & $4.60\sigma$ \tabularnewline
{ } 
Case I +R21  & $0.02302\pm0.00021$  & $0.079_{-0.019}^{+0.021}$  & $70.85\pm0.77$  & $1.89\sigma$ \tabularnewline
{ } 
Case II & $0.02245_{-0.00015}^{+0.00013}$  & $0.0165_{-0.016}^{+0.0034}$  & $67.82_{-0.54}^{+0.42}$  & $4.78\sigma$ \tabularnewline
{ } 
Case III & $0.02251_{-0.00016}^{+0.00013}$  & $0.0177_{-0.018}^{+0.0035}$  & $67.99_{-0.62}^{+0.38}$  & $4.58\sigma$ \tabularnewline
{ } 
Case III+R21 & $0.02293\pm0.00019$  & $0.075_{-0.018}^{+0.022}$  & $70.65\pm0.76$  & $2.06 \sigma$ \tabularnewline
{ } 
Case IV  & $0.02232\pm0.00012$  & $0.0131_{-0.013}^{+0.00226}$  & $67.52\pm0.42$  & $5.15\sigma$ \tabularnewline
{ } 
Case V  & $0.02227\pm0.00011$  & $0.0135_{-0.014}^{+0.0026}$  & $67.44_{-0.45}^{+0.38}$  & $5.23\sigma$ \tabularnewline
{ } 
Case V +R21 & $0.02243\pm0.00015$  & $0.044_{-0.020}^{+0.027}$  & $69.31_{-0.80}^{+0.92}$  & $2.95\sigma$ \tabularnewline
{ } 
Case VI  & $0.02246_{-0.00015}^{+0.00013}$  & $0.0174_{-0.017}^{+0.0038}$  & $67.87_{-0.57}^{+0.40}$  & $4.72\sigma$ \tabularnewline
{ } 
Case VII  & $0.02243\pm0.00014$  & $0.0149_{-0.015}^{+0.0029}$  & $67.76_{-0.48}^{+0.41}$  & $4.90\sigma$ \tabularnewline
\hline \hline
\end{tabular}\caption{\label{tab:EDE_ninf} Derived values of $\Omega_{b}h^{2},f_{{\rm de}}(z_{c})$, $H_{0}$ and the Gaussian tension for the DE model with $n=\infty$. }
\end{table}

\pagebreak
\section{Conclusion and discussion \label{sec:conclusion}}

We have investigated the impact of the BBN on the Hubble tension,  focusing on to what extent the evaluation of the tension is affected by how we include the BBN in the analysis.  In models proposed to resolve the $H_0$ tension, not only the derived value of $H_0$ tends to be increased in the fitting to cosmological data such as CMB, the determination of other cosmological parameters is also affected, among which the baryon density is significantly affected.  Indeed the baryon density can also be well determined by the BBN data, in particular the primordial deuterium abundance.  We have investigated this issue in the $\Lambda$CDM model,  and the EDE as an example of models which has been proposed to resolve the $H_0$ tension.

When one includes the BBN data in addition to CMB and BAO in the analysis, there are several ways to take account of it  as to which observational data is used and which nuclear reaction rate, especially for  $d (p, \gamma) ^3 {\rm He}$ which is particularly relevant to the deuterium abundance, is adopted. We have made the analysis adoping two different observational data for $D_p$ and three separate treatments for the reaction rate of $d (p, \gamma) ^3 {\rm He}$. We found that the significance of the $H_0$ tension  between the direct measurement of Cepheid calibrated supernova distance ladder from \cite{Riess:2021jrx} and indirect ones from Planck+BAO in combination with BBN data varies depending on which observational data and the nuclear reaction rate are used in the analysis. 

In $\Lambda$CDM model, the tension is $5.04\sigma$ for Case~II where the value of $D_p$ from Cooke et al. \cite{Cooke:2017cwo} and the reaction rate of Adelberger et al. \cite{Adelberger:2010qa}  for $d (p, \gamma) ^3 {\rm He}$  are assumed, which is usually adopted in the literature. However, it increases to  $5.38\sigma$ for Case~V where the weighted average of 11 recent measurements of $D_p$  \cite{ParticleDataGroup:2022pth} and the reaction rate from Marcucci et al.  \cite{Marcucci:2015yla} are assumed. 

The impact is more significant in the EDE model, which has been extensively studied as a possible solution to the $H_0$ tension. For the case of the EDE model with $n=2$, the significance is increased from $3.83\sigma$ (Case~II) to $4.41\sigma$ (Case~V), which indicates that the assumption in the treatment of BBN would significantly affect the tension. Our analysis shows that we need to carefully consider which data and the reaction rate are adopted in the BBN analysis to investigate the Hubble tension.

Since the Hubble tension is now one of the most serious problem in cosmology today, one needs to carefully quantify the tension particularly when we analyze models to resolve the $H_0$ tension.  We  have various accurate cosmological observations such as CMB,  BAO and BBN and the combinations of these data would constrain cosmological models even further. However, we may have some options as to which data and how the theoretical calculation are performed, which can appreciably affect the significance of  the $H_0$ tension. We have explicitly demonstrated this issue for the BBN, which gives some cautions to future research in the $H_0$ tension.

\section*{Acknowledgements}
This work was supported in part by JSPS KAKENHI Grant Number 19K03874~(TT), MEXT KAKENHI Grant Number  23H04515~(TT), and JST SPRING, Grant Number JPMJSP2119~(YT).

\clearpage 

\bibliography{BBN_H0}

\providecommand{\href}[2]{#2}\begingroup\raggedright\begin{thebibliography}{10}

\bibitem{Riess:2021jrx}
A.~G. Riess et~al., {\it {A Comprehensive Measurement of the Local Value of the
  Hubble Constant with 1 km s$^{-1}$ Mpc$^{-1}$ Uncertainty from the Hubble
  Space Telescope and the SH0ES Team}},  {\em Astrophys. J. Lett.} {\bf 934}
  (2022), no.~1 L7, [\href{http://arxiv.org/abs/2112.04510}{{\tt
  arXiv:2112.04510}}].

\bibitem{Planck:2018vyg}
{\bf Planck} Collaboration, N.~Aghanim et~al., {\it {Planck 2018 results. VI.
  Cosmological parameters}},  {\em Astron. Astrophys.} {\bf 641} (2020) A6,
  [\href{http://arxiv.org/abs/1807.06209}{{\tt arXiv:1807.06209}}]. [Erratum:
  Astron.Astrophys. 652, C4 (2021)].

\bibitem{Schoneberg:2019wmt}
N.~Sch\"oneberg, J.~Lesgourgues, and D.~C. Hooper, {\it {The BAO+BBN take on
  the Hubble tension}},  {\em JCAP} {\bf 10} (2019) 029,
  [\href{http://arxiv.org/abs/1907.11594}{{\tt arXiv:1907.11594}}].

\bibitem{Okamatsu:2021jil}
F.~Okamatsu, T.~Sekiguchi, and T.~Takahashi, {\it {$H_0$ tension without CMB
  data: Beyond the \ensuremath{\Lambda}CDM}},  {\em Phys. Rev. D} {\bf 104}
  (2021), no.~2 023523, [\href{http://arxiv.org/abs/2105.12312}{{\tt
  arXiv:2105.12312}}].

\bibitem{Schoneberg:2022ggi}
N.~Sch\"oneberg, L.~Verde, H.~Gil-Mar\'\i{}n, and S.~Brieden, {\it {BAO+BBN
  revisited -- Growing the Hubble tension with a 0.7km/s/Mpc constraint}},
  \href{http://arxiv.org/abs/2209.14330}{{\tt arXiv:2209.14330}}.

\bibitem{DiValentino:2021izs}
E.~Di~Valentino, O.~Mena, S.~Pan, L.~Visinelli, W.~Yang, A.~Melchiorri, D.~F.
  Mota, A.~G. Riess, and J.~Silk, {\it {In the realm of the Hubble
  tension\textemdash{}a review of solutions}},  {\em Class. Quant. Grav.} {\bf
  38} (2021), no.~15 153001, [\href{http://arxiv.org/abs/2103.01183}{{\tt
  arXiv:2103.01183}}].

\bibitem{Perivolaropoulos:2021jda}
L.~Perivolaropoulos and F.~Skara, {\it {Challenges for \ensuremath{\Lambda}CDM:
  An update}},  {\em New Astron. Rev.} {\bf 95} (2022) 101659,
  [\href{http://arxiv.org/abs/2105.05208}{{\tt arXiv:2105.05208}}].

\bibitem{Schoneberg:2021qvd}
N.~Sch\"oneberg, G.~Franco~Abell\'an, A.~P\'erez~S\'anchez, S.~J. Witte,
  V.~Poulin, and J.~Lesgourgues, {\it {The H0 Olympics: A fair ranking of
  proposed models}},  {\em Phys. Rept.} {\bf 984} (2022) 1--55,
  [\href{http://arxiv.org/abs/2107.10291}{{\tt arXiv:2107.10291}}].

\bibitem{Ichikawa:2004zi}
K.~Ichikawa, M.~Fukugita, and M.~Kawasaki, {\it {Constraining neutrino masses
  by CMB experiments alone}},  {\em Phys. Rev. D} {\bf 71} (2005) 043001,
  [\href{http://arxiv.org/abs/astro-ph/0409768}{{\tt astro-ph/0409768}}].

\bibitem{Ichikawa:2008pz}
K.~Ichikawa, T.~Sekiguchi, and T.~Takahashi, {\it {Probing the Effective Number
  of Neutrino Species with Cosmic Microwave Background}},  {\em Phys. Rev. D}
  {\bf 78} (2008) 083526, [\href{http://arxiv.org/abs/0803.0889}{{\tt
  arXiv:0803.0889}}].

\bibitem{Sekiguchi:2009zs}
T.~Sekiguchi, K.~Ichikawa, T.~Takahashi, and L.~Greenhill, {\it {Neutrino mass
  from cosmology: Impact of high-accuracy measurement of the Hubble constant}},
   {\em JCAP} {\bf 03} (2010) 015, [\href{http://arxiv.org/abs/0911.0976}{{\tt
  arXiv:0911.0976}}].

\bibitem{Takahashi:2022cpn}
T.~Takahashi and S.~Yamashita, {\it {Big bang nucleosynthesis and early dark
  energy in light of the EMPRESS Yp results and the H0 tension}},  {\em Phys.
  Rev. D} {\bf 107} (2023), no.~10 103520,
  [\href{http://arxiv.org/abs/2211.04087}{{\tt arXiv:2211.04087}}].

\bibitem{Seto:2021xua}
O.~Seto and Y.~Toda, {\it {Comparing early dark energy and extra radiation
  solutions to the Hubble tension with BBN}},  {\em Phys. Rev. D} {\bf 103}
  (2021), no.~12 123501, [\href{http://arxiv.org/abs/2101.03740}{{\tt
  arXiv:2101.03740}}].

\bibitem{Seto:2022xgx}
O.~Seto and Y.~Toda, {\it {Big bang nucleosynthesis constraints on varying
  electron mass solution to the Hubble tension}},  {\em Phys. Rev. D} {\bf 107}
  (2023), no.~8 083512, [\href{http://arxiv.org/abs/2206.13209}{{\tt
  arXiv:2206.13209}}].

\bibitem{Seto:2021tad}
O.~Seto and Y.~Toda, {\it {Hubble tension in lepton asymmetric cosmology with
  an extra radiation}},  {\em Phys. Rev. D} {\bf 104} (2021), no.~6 063019,
  [\href{http://arxiv.org/abs/2104.04381}{{\tt arXiv:2104.04381}}].

\bibitem{Sekiguchi:2020igz}
T.~Sekiguchi and T.~Takahashi, {\it {Cosmological bound on neutrino masses in
  the light of $H_0$ tension}},  {\em Phys. Rev. D} {\bf 103} (2021), no.~8
  083516, [\href{http://arxiv.org/abs/2011.14481}{{\tt arXiv:2011.14481}}].

\bibitem{Ye:2021nej}
G.~Ye, B.~Hu, and Y.-S. Piao, {\it {Implication of the Hubble tension for the
  primordial Universe in light of recent cosmological data}},  {\em Phys. Rev.
  D} {\bf 104} (2021), no.~6 063510,
  [\href{http://arxiv.org/abs/2103.09729}{{\tt arXiv:2103.09729}}].

\bibitem{Takahashi:2021bti}
F.~Takahashi and W.~Yin, {\it {Cosmological implications of $n_s \approx 1$ in
  light of the Hubble tension}},  {\em Phys. Lett. B} {\bf 830} (2022) 137143,
  [\href{http://arxiv.org/abs/2112.06710}{{\tt arXiv:2112.06710}}].

\bibitem{Jiang:2022uyg}
J.-Q. Jiang and Y.-S. Piao, {\it {Toward early dark energy and ns=1 with
  Planck, ACT, and SPT observations}},  {\em Phys. Rev. D} {\bf 105} (2022),
  no.~10 103514, [\href{http://arxiv.org/abs/2202.13379}{{\tt
  arXiv:2202.13379}}].

\bibitem{Ye:2022efx}
G.~Ye, J.-Q. Jiang, and Y.-S. Piao, {\it {Towards hybrid inflation with $n_s=1$
  in light of Hubble tension and primordial gravitational waves}},
  \href{http://arxiv.org/abs/2205.02478}{{\tt arXiv:2205.02478}}.

\bibitem{Sekiguchi:2020teg}
T.~Sekiguchi and T.~Takahashi, {\it {Early recombination as a solution to the
  $H_0$ tension}},  {\em Phys. Rev. D} {\bf 103} (2021), no.~8 083507,
  [\href{http://arxiv.org/abs/2007.03381}{{\tt arXiv:2007.03381}}].

\bibitem{Poulin:2018cxd}
V.~Poulin, T.~L. Smith, T.~Karwal, and M.~Kamionkowski, {\it {Early Dark Energy
  Can Resolve The Hubble Tension}},  {\em Phys. Rev. Lett.} {\bf 122} (2019),
  no.~22 221301, [\href{http://arxiv.org/abs/1811.04083}{{\tt
  arXiv:1811.04083}}].

\bibitem{Aver:2015iza}
E.~Aver, K.~A. Olive, and E.~D. Skillman, {\it {The effects of He I
  \ensuremath{\lambda}10830 on helium abundance determinations}},  {\em JCAP}
  {\bf 07} (2015) 011, [\href{http://arxiv.org/abs/1503.08146}{{\tt
  arXiv:1503.08146}}].

\bibitem{Hsyu:2020uqb}
T.~Hsyu, R.~J. Cooke, J.~X. Prochaska, and M.~Bolte, {\it {The PHLEK Survey: A
  New Determination of the Primordial Helium Abundance}},  {\em Astrophys. J.}
  {\bf 896} (2020), no.~1 77, [\href{http://arxiv.org/abs/2005.12290}{{\tt
  arXiv:2005.12290}}].

\bibitem{Kurichin:2021ppm}
O.~A. Kurichin, P.~A. Kislitsyn, V.~V. Klimenko, S.~A. Balashev, and A.~V.
  Ivanchik, {\it {A new determination of the primordial helium abundance using
  the analyses of H II region spectra from SDSS}},  {\em Mon. Not. Roy. Astron.
  Soc.} {\bf 502} (2021), no.~2 3045--3056,
  [\href{http://arxiv.org/abs/2101.09127}{{\tt arXiv:2101.09127}}].

\bibitem{Matsumoto:2022tlr}
A.~Matsumoto et~al., {\it {EMPRESS. VIII. A New Determination of Primordial He
  Abundance with Extremely Metal-poor Galaxies: A Suggestion of the Lepton
  Asymmetry and Implications for the Hubble Tension}},  {\em Astrophys. J.}
  {\bf 941} (2022), no.~2 167, [\href{http://arxiv.org/abs/2203.09617}{{\tt
  arXiv:2203.09617}}].

\bibitem{Ichikawa:2006dt}
K.~Ichikawa and T.~Takahashi, {\it {Revisiting the constraint on the helium
  abundance from cmb}},  {\em Phys. Rev. D} {\bf 73} (2006) 063528,
  [\href{http://arxiv.org/abs/astro-ph/0601099}{{\tt astro-ph/0601099}}].

\bibitem{Ichikawa:2007js}
K.~Ichikawa, T.~Sekiguchi, and T.~Takahashi, {\it {Primordial Helium Abundance
  from CMB: a constraint from recent observations and a forecast}},  {\em Phys.
  Rev. D} {\bf 78} (2008) 043509, [\href{http://arxiv.org/abs/0712.4327}{{\tt
  arXiv:0712.4327}}].

\bibitem{Cooke:2017cwo}
R.~J. Cooke, M.~Pettini, and C.~C. Steidel, {\it {One Percent Determination of
  the Primordial Deuterium Abundance}},  {\em Astrophys. J.} {\bf 855} (2018),
  no.~2 102, [\href{http://arxiv.org/abs/1710.11129}{{\tt arXiv:1710.11129}}].

\bibitem{ParticleDataGroup:2022pth}
{\bf Particle Data Group} Collaboration, R.~L. Workman et~al., {\it {Review of
  Particle Physics}},  {\em PTEP} {\bf 2022} (2022) 083C01.

\bibitem{Adelberger:2010qa}
E.~G. Adelberger et~al., {\it {Solar fusion cross sections II: the pp chain and
  CNO cycles}},  {\em Rev. Mod. Phys.} {\bf 83} (2011) 195,
  [\href{http://arxiv.org/abs/1004.2318}{{\tt arXiv:1004.2318}}].

\bibitem{Pisanti:2020efz}
O.~Pisanti, G.~Mangano, G.~Miele, and P.~Mazzella, {\it {Primordial Deuterium
  after LUNA: concordances and error budget}},  {\em JCAP} {\bf 04} (2021) 020,
  [\href{http://arxiv.org/abs/2011.11537}{{\tt arXiv:2011.11537}}].

\bibitem{Marcucci:2015yla}
L.~E. Marcucci, G.~Mangano, A.~Kievsky, and M.~Viviani, {\it {Implication of
  the proton-deuteron radiative capture for Big Bang Nucleosynthesis}},  {\em
  Phys. Rev. Lett.} {\bf 116} (2016), no.~10 102501,
  [\href{http://arxiv.org/abs/1510.07877}{{\tt arXiv:1510.07877}}]. [Erratum:
  Phys.Rev.Lett. 117, 049901 (2016)].

\bibitem{Poulin:2018dzj}
V.~Poulin, T.~L. Smith, D.~Grin, T.~Karwal, and M.~Kamionkowski, {\it
  {Cosmological implications of ultralight axionlike fields}},  {\em Phys. Rev.
  D} {\bf 98} (2018), no.~8 083525,
  [\href{http://arxiv.org/abs/1806.10608}{{\tt arXiv:1806.10608}}].

\bibitem{Braglia:2020bym}
M.~Braglia, W.~T. Emond, F.~Finelli, A.~E. Gumrukcuoglu, and K.~Koyama, {\it
  {Unified framework for early dark energy from $\alpha$-attractors}},  {\em
  Phys. Rev. D} {\bf 102} (2020), no.~8 083513,
  [\href{http://arxiv.org/abs/2005.14053}{{\tt arXiv:2005.14053}}].

\bibitem{Agrawal:2019lmo}
P.~Agrawal, F.-Y. Cyr-Racine, D.~Pinner, and L.~Randall, {\it {Rock 'n' Roll
  Solutions to the Hubble Tension}},
  \href{http://arxiv.org/abs/1904.01016}{{\tt arXiv:1904.01016}}.

\bibitem{Kamionkowski:2022pkx}
M.~Kamionkowski and A.~G. Riess, {\it {The Hubble Tension and Early Dark
  Energy}},  \href{http://arxiv.org/abs/2211.04492}{{\tt arXiv:2211.04492}}.

\bibitem{Poulin:2023lkg}
V.~Poulin, T.~L. Smith, and T.~Karwal, {\it {The Ups and Downs of Early Dark
  Energy solutions to the Hubble tension: a review of models, hints and
  constraints circa 2023}},  \href{http://arxiv.org/abs/2302.09032}{{\tt
  arXiv:2302.09032}}.

\bibitem{Minami:2020odp}
Y.~Minami and E.~Komatsu, {\it {New Extraction of the Cosmic Birefringence from
  the Planck 2018 Polarization Data}},  {\em Phys. Rev. Lett.} {\bf 125}
  (2020), no.~22 221301, [\href{http://arxiv.org/abs/2011.11254}{{\tt
  arXiv:2011.11254}}].

\bibitem{Diego-Palazuelos:2022dsq}
P.~Diego-Palazuelos et~al., {\it {Cosmic Birefringence from the Planck Data
  Release 4}},  {\em Phys. Rev. Lett.} {\bf 128} (2022), no.~9 091302,
  [\href{http://arxiv.org/abs/2201.07682}{{\tt arXiv:2201.07682}}].

\bibitem{Capparelli:2019rtn}
L.~M. Capparelli, R.~R. Caldwell, and A.~Melchiorri, {\it {Cosmic birefringence
  test of the Hubble tension}},  {\em Phys. Rev. D} {\bf 101} (2020), no.~12
  123529, [\href{http://arxiv.org/abs/1909.04621}{{\tt arXiv:1909.04621}}].

\bibitem{Murai:2022zur}
K.~Murai, F.~Naokawa, T.~Namikawa, and E.~Komatsu, {\it {Isotropic cosmic
  birefringence from early dark energy}},  {\em Phys. Rev. D} {\bf 107} (2023),
  no.~4 L041302, [\href{http://arxiv.org/abs/2209.07804}{{\tt
  arXiv:2209.07804}}].

\bibitem{camb}
\url{http://camb.info}.

\bibitem{Lewis:2002ah}
A.~Lewis and S.~Bridle, {\it {Cosmological parameters from CMB and other data:
  A Monte Carlo approach}},  {\em Phys. Rev. D} {\bf 66} (2002) 103511,
  [\href{http://arxiv.org/abs/astro-ph/0205436}{{\tt astro-ph/0205436}}].

\bibitem{Planck:2019nip}
{\bf Planck} Collaboration, N.~Aghanim et~al., {\it {Planck 2018 results. V.
  CMB power spectra and likelihoods}},  {\em Astron. Astrophys.} {\bf 641}
  (2020) A5, [\href{http://arxiv.org/abs/1907.12875}{{\tt arXiv:1907.12875}}].

\bibitem{Beutler:2011hx}
F.~Beutler, C.~Blake, M.~Colless, D.~H. Jones, L.~Staveley-Smith, L.~Campbell,
  Q.~Parker, W.~Saunders, and F.~Watson, {\it {The 6dF Galaxy Survey: Baryon
  Acoustic Oscillations and the Local Hubble Constant}},  {\em Mon. Not. Roy.
  Astron. Soc.} {\bf 416} (2011) 3017--3032,
  [\href{http://arxiv.org/abs/1106.3366}{{\tt arXiv:1106.3366}}].

\bibitem{Ross:2014qpa}
A.~J. Ross, L.~Samushia, C.~Howlett, W.~J. Percival, A.~Burden, and M.~Manera,
  {\it {The clustering of the SDSS DR7 main Galaxy sample \textendash{} I. A 4
  per cent distance measure at $z = 0.15$}},  {\em Mon. Not. Roy. Astron. Soc.}
  {\bf 449} (2015), no.~1 835--847, [\href{http://arxiv.org/abs/1409.3242}{{\tt
  arXiv:1409.3242}}].

\bibitem{BOSS:2016wmc}
{\bf BOSS} Collaboration, S.~Alam et~al., {\it {The clustering of galaxies in
  the completed SDSS-III Baryon Oscillation Spectroscopic Survey: cosmological
  analysis of the DR12 galaxy sample}},  {\em Mon. Not. Roy. Astron. Soc.} {\bf
  470} (2017), no.~3 2617--2652, [\href{http://arxiv.org/abs/1607.03155}{{\tt
  arXiv:1607.03155}}].

\bibitem{Consiglio:2017pot}
R.~Consiglio, P.~F. de~Salas, G.~Mangano, G.~Miele, S.~Pastor, and O.~Pisanti,
  {\it {PArthENoPE reloaded}},  {\em Comput. Phys. Commun.} {\bf 233} (2018)
  237--242, [\href{http://arxiv.org/abs/1712.04378}{{\tt arXiv:1712.04378}}].

\bibitem{Gariazzo:2021iiu}
S.~Gariazzo, P.~F.~de Salas, O.~Pisanti, and R.~Consiglio, {\it {PArthENoPE
  revolutions}},  {\em Comput. Phys. Commun.} {\bf 271} (2022) 108205,
  [\href{http://arxiv.org/abs/2103.05027}{{\tt arXiv:2103.05027}}].

\end{thebibliography}\endgroup

\end{document}